\input ./style/arxiv-general.cfg
\documentclass[aoas,MSNbibl,nameyear,noautosecdot,dvips]{arximspdf}
\makeatletter
   \@ifpackageloaded{graphicx}{}{\usepackage{graphicx}}
\makeatother
\usepackage{algorithm}
\usepackage{url,breakurl}

\doi{10.1214/15-AOAS831}
\volume{9}
\issue{3}
\pubyear{2015}
\firstpage{1394}
\lastpage{1414}
\docsubty{FLA}

\begin{document}
\begin{frontmatter}

\title{Distributed multinomial regression}
\runtitle{Distributed multinomial regression}

\begin{aug}
\author[A]{\fnms{Matt}~\snm{Taddy}\corref{}\ead[label=e1]{taddy@chicagobooth.edu}}
\runauthor{M. Taddy}
\affiliation{University of Chicago}
\address[A]{University of Chicago\\
\quad Booth School of Business\\
5807 South Woodlawn Avenue\\
Chicago,  Illinois 60637\\
USA\\
\printead{e1}}
\end{aug}

%
\received{\smonth{12} \syear{2013}}
%
\revised{\smonth{4} \syear{2015}}

%
\begin{abstract}
This article introduces a model-based approach to distributed
computing for multinomial logistic (softmax) regression. We treat
counts for
each response category as independent Poisson regressions via plug-in
estimates for fixed effects shared across categories. The work is
driven by
the high-dimensional-response multinomial models that are used in
analysis of
a large number of random counts. Our motivating applications are in text
analysis, where documents are tokenized and the token counts are
modeled as
arising from a multinomial dependent upon document attributes. We estimate
such models for a publicly available data set of reviews from Yelp,
with text
regressed onto a large set of explanatory variables (user, business, and
rating information). The fitted models serve as a basis for exploring the
connection between words and variables of interest, for reducing dimension
into supervised factor scores, and for prediction. We argue that the approach
herein provides an attractive option for social scientists and other text
analysts who wish to bring familiar regression tools to bear on text data.
\end{abstract}

%
\begin{keyword}
\kwd{Distributed computing}
\kwd{MapReduce}
\kwd{logistic regression}
\kwd{lasso}
\kwd{text analysis}
\kwd{multinomial inverse regression}
\kwd{computational social science}
\end{keyword}
\end{frontmatter}

\section{\texorpdfstring{Introduction.}{Introduction}}
\label{intro}

This article is motivated by data sets that include\break \textit{counts} in
a massive
number of \textit{categories}, such as text corpora (counts for
words), browser
logs (counts on websites), and website tracking (counts of clicks). The unit
upon which counts are observed---for example, a ``document'' for text
or a ``user'' in
web analysis---is annotated with \textit{attributes}, additional information
about each document (author, date, etc.) or user (age, purchases,
etc.). Much
of contemporary Big Data analysis involves some exploration, inference, and
prediction of, or controlling for, the relationship between these attributes
and the associated very high-dimensional counts.

Say $\mathbf{c}_i$ is a vector of counts in $d$ categories, summing to
$m_i =
\sum_j c_{ij}$, accompanied by a $p$-dimensional attribute vector
$\mathbf{v}_i$ on observation
unit $i$ of $n$ total. For example, in the archetypal text mining
application, $\mathbf{c}_i$
are counts for words in document $i$ annotated with metadata $\mathbf
{v}_i$. We
connect attributes and counts through a big multinomial logistic regression
model,
\begin{eqnarray}
 && \mathrm{p}(\mathbf{c}_{i} | \mathbf{v}_i,m_i)
= \operatorname{MN} (\mathbf{c}_i; \mathbf{q}_{i}, m_i )
\nonumber
\\[-8pt]
\label{bigmn}
\\[-8pt]
\eqntext{\displaystyle \mbox{where } q_{ij} = \frac{e^{\eta_{ij}}}{\Lambda
_i},\eta_{ij} =
\alpha_j + \mathbf{v}_i'\bolds{\varphi}_j \quad\mbox{and}\quad
\Lambda_i = \sum
_{k=1}^d e^{\eta_{ik}}.}
\end{eqnarray}
The multinomial denoted $\mathrm{MN}$ here has, for unit $i$, category $j$,
probability $q_{ij}$, and size $m_i$. This model can be computationally
expensive to estimate for a large number of response categories (i.e., big
$\mathbf{c}_i$ dimension $d$). Even a single likelihood evaluation is
costly, due
to the sum required for each normalizing \textit{intensity} $\Lambda
_i =
\sum_{k=1}^d e^{\eta_{ik}}$. The methodological innovation of the current
article is to replace $\Lambda_i$ with initial estimates, then
condition upon
these plug-ins when estimating (\ref{bigmn}) through $d$ individual Poisson
regressions for counts in each category $j$. This model-based factorization
allows one to {\it partition} computation across many independent
machines, so
with enough processors the system of (\ref{bigmn}) is fit in the time required
for a single Poisson regression.

We refer to this framework as \textit{distributed multinomial regression},
or DMR.
Our work here extends ideas from \citet{taddymultinomial2013}, which
introduced the strategy of \textit{multinomial inverse regression}
(MNIR). That
article argues for estimation of models like (\ref{bigmn}) as the
first step
in an inverse regression routine for predicting elements of new
$\mathbf{v}_i$.
However, \citet{taddymultinomial2013} relies upon a fitting
algorithm that
collapses response counts across equal $\mathbf{v}_i$, and hence scales
{\it only}
for a small number of attributes (i.e., when $p$ is just one or two). That
article is also focused exclusively on applications in attribute prediction.
The purpose of the current article is thus twofold: to supply
techniques for
estimation when both $\mathbf{c}$ and $\mathbf{v}$ are high
dimensional, and to
motivate how these models can be useful in many aspects of analysis and
inference.

Much of the paper is devoted to an example analysis of reviews on Yelp---an
Internet platform for feedback on various establishments, including
restaurants, barbers, schools, and much else. This data set has a rich feature
set associated with a wide variety of reviews. The data are also publicly
available, after (free) registration on the data mining contest website
\url{kaggle.com}.
Moreover, our technology is provided in the {\tt distrom} package for {\tt R} and
Yelp analysis code is cataloged at \url{github.com/mataddy/yelp}. Public
access is essential here: our goal is to provide a
complete template for analysis of high-dimensional count data.

The estimation strategy is detailed in Section~\ref{methods},
including model
factorization, plug-ins for $\Lambda_i$, and regularization
path estimation within each parallel regression. Methods
are illustrated in the short classification example of Section~\ref
{FGL}, which shows utility for DMR not only in
big $d$ but also as a speedup for small $d$ multinomial regressions. Finally,
Section~\ref{YELP} runs through our full Yelp application, detailing model
estimation and a variety of analyses. These analyses each correspond to a
different inferential goal.

\subsection*{\texorpdfstring{Exploration: What words are associated with funny or useful
content?}{Exploration: What words are associated with funny or useful
content?}}
Here,
we interpret the fitted regression model at the level of word loadings.
We emphasize that these loadings represent \textit{partial
effects}---connections
between text and attributes that arise after controlling for
collinearity between attributes---and we describe how the
interpretations change when controlling for more or less confounding
information.

\subsection*{\texorpdfstring{Dimension reduction: Which reviews have the most funny or useful
content?}{Dimension reduction: Which reviews have the most funny or useful
content?}} Once the model has been fit, it acts as a map between text and
attributes. We describe how to use this map to obtain \textit{sufficient
reductions}: low-dimensional scores that summarize text content \textit
{directly}
relevant to a given attribute. We show that the sufficient reduction for,
say, funny votes, is a seemingly better judge of humor than the
original yelp
voters (because reviews can generate funny votes for correlated but unfunny
reasons).

\subsection*{\texorpdfstring{Prediction: What will be the usefulness or hilarity of a new
review?}{Prediction: What will be the usefulness or hilarity of a new
review?}} We consider performance of the above sufficient reductions as
input to prediction algorithms, and find that they can match or
outperform comparable techniques that use the full text as input.

\subsection*{\texorpdfstring{Treatment effects: Does user tenure lead to higher ratings?}{Treatment effects: Does user tenure lead to higher ratings?}}
Finally, we
show how to use the sufficient reductions as synthetic controls in regressions
that wish to remove from a targeted treatment effect the influence of any
correlated text content. That is, the sufficient reductions act like a
text-based propensity; in our application, we use them to see if older
users are
more positive in their ratings even if we control for the change in their
review content.

Section~\ref{END} closes with a discussion and some practical advice for
applications in text mining. We argue that, for many social science and
business applications, the methods herein make regression of text onto
observed attributes an attractive alternative to techniques such as topic
modeling which require estimation and interpretation of a latent space.

\section{\texorpdfstring{Methods: Estimation in distribution.}{Methods: Estimation in distribution}}
\label{methods}

We adopt terminology from text analysis for the remainder
and refer to each unit $i$ as a ``document'' and each category $j$ as a
``word.''\footnote{Even in text mining this is a simplification; each
$j$ could
be a combination of words or any other language token.} Suppose that every
document--word count $c_{ij}$ has been drawn independently
$\operatorname{Po} (e^{\eta_{ij}} )$---Poisson with intensity (i.e., mean)
$e^{\eta_{ij}}$. The joint document likelihood for $\mathbf{c}_i$ then
factorizes as the
product of a multinomial distribution for individual counts conditional on
total count $m_i$ and a Poisson distribution on~$m_i$:
\begin{equation}
\label{embed}
\mathrm{p}(\mathbf{c}_{i}) = \prod_j
\operatorname{Po} \bigl(c_{ij}; e^{\eta_{ij}} \bigr) = \operatorname{MN} (
\mathbf{c}_i; \mathbf{q}_i, m_i )\operatorname{Po}
(m_i;\Lambda_i ).
\end{equation}

This well-known result has long been used by statisticians to justify ignoring
whether sampling was conditional on margin totals in analysis of contingency
tables. \citet{birchmaximum1963} showed that the maximum
likelihood estimate
(MLE) of $\mathbf{q}_i$ is unchanged under a variety of sampling
models for
3-way tables {\it under the constraint} that $\Lambda_{i} = m_i$. This is
satisfied at the MLE for a saturated model. \citet{palmgrenfisher1981}
extends the theory to log-linear regression with $\eta_{ij} = \alpha_j
+ \mu_i + \bolds{\varphi}_j'\mathbf{v}_i$, showing that the Fisher
information on
coefficients is the same regardless of whether or not you have
conditioned on $m_i$ so long as $\mu_i$ in the Poisson model is estimated
at its conditional MLE,
\begin{equation}
\label{mlemu} \mu_i^\star= \log\biggl(\frac{m_i}{\sum_j e^{\eta_{ij}}}
\biggr).
\end{equation}

Most commonly, (\ref{embed}) is invoked when applying multinomial logistic
regression: totals $m_i$ are then ancillary and the $\mu_i$ drop out
of the
likelihood. Our DMR framework takes the opposite view: if we are
willing to
fix estimates $\hat\mu_i$ potentially not at their MLE (we will argue for
$\hat\mu_i = \log m_i$), then the factorized Poisson likelihood can be
analyzed independently across response categories.\footnote{In an
older version
of this idea, \citet{hodgespoisson1960} introduce a Poisson approximation
to the binomial distribution, for which
\citet{mcdonaldpoisson1980} provides error bounds and extension to
multinomials.} As highlighted in the \hyperref[intro]{Introduction}, this yields distributed
computing algorithms for estimation on previously impossible scales. Indeed,
we have observed in text and web analysis a recent migration from multinomial
models---say, for latent factorization---to Poisson model schemes; see
\citet{gopalanscalable2013} as an example. From the perspective
of this
article, such strategies are Big Data approximations to their
multinomial precursors.

\subsection{\texorpdfstring{Estimating baseline intensity.}{Estimating baseline intensity}}
\label{MU}

The negative log likelihood implied by (\ref{bigmn}) is proportional to
\begin{equation}
\label{mnl} \sum_{i=1}^n \Biggl[
m_i\log\Biggl(\sum_{j=1}^d
e^{\eta
_{ij}} \Biggr) - \mathbf{c}_{i}'\bolds{\eta}_{i} \Biggr].
\end{equation}
It is easy to verify that adding observation fixed effects $\mu_i$ to each
$\eta_{ij}$ in (\ref{mnl}) leaves the likelihood unchanged. In
contrast, the
corresponding Poisson model, unconditional on $m_i$, has negative log
likelihood proportional to
\begin{equation}
\label{pol} \sum_{j=1}^d\sum
_{i=1}^n \bigl[ e^{\mu_i + \eta_{ij}} - c_{ij}(
\mu_i + \eta_{ij}) \bigr]
\end{equation}
with gradient on each $\mu_i$ of $g(\mu_i) =
e^{\mu_i}\sum_j e^{\eta_{ij}} - m_i$, and is clearly sensitive to these
observation ``baseline intensities.'' As mentioned above, the solution
for the
parameters of $\eta_{ij}$ is unchanged between (\ref{mnl}) and (\ref
{pol}) if
each $\mu_i$ is set to its conditional MLE in (\ref{mlemu}).

Unfortunately, if our goal is to \textit{separate} inference for
$\bolds{\varphi}_j$
across different $j$, the MLE formula of (\ref{mlemu}) will create a
computational bottleneck: each category-$j$ Poisson regression requires
updates to $\bolds{\mu}^\star= [\mu^\star_1 \cdots\mu^\star
_n]'$ during
estimation. Distributed computation precludes such communication, and
we instead use the simple plug-in estimator
\begin{equation}
\label{plugin} \hat\mu_i = \log m_i.
\end{equation}
This choice is justified as optimal in a few simple models, and we rely upon
empirical evidence to claim it performs well in more complex
settings.\footnote{Note that, when compared to (\ref{mlemu}), the plug-in
replaces $\sum_j e^{\alpha_j
+ \mathbf{v}_i'\bolds{\varphi}_j}$ with 1. Adding a constant to each
$\alpha_j$ leaves probabilities unchanged, so this can be made to hold without
affecting fit.}

The gradient of the Poisson likelihood in (\ref{pol}) on $\mu_i$ at our
plug-in is $g(\hat\mu_i) = m_i (\sum_i e^{\eta_{ij}}-1 )$.
Define the plug-in MLEs $\hat{\bolds{\eta}}_{i}
= [\hat\eta_{i1}\cdots\hat\eta_{id}]'$ as those which minimize the Poisson
objective in (\ref{pol}) under $\mu_i=\hat\mu_i$. Then in the
three simple settings below, $g(\hat
\mu_i)=0$ for $\bolds{\eta}_i = \hat{\bolds{\eta}}_{i}$. This
implies that $\hat
\mu_i$
is the optimal joint MLE estimate for $\mu_i$, and thus that
$\{\hat{\bolds{\eta}}_{i},\hat\mu_i\}$ minimize the Poisson
objective in
(\ref{pol}) while $\{ \hat{\bolds{\eta}}_{i}\}$ minimizes the
logistic multinomial
objective in (\ref{mnl}):
\begin{itemize}
\item In a saturated model, with
each $\eta_{ij}$ free, $\hat
\eta_{ij} = \log(c_{ij}) - \hat\mu_i = \log(c_{ij}/m_i)$ and
$g(\hat
\mu_i) = 0$.
\item With intercept-only $\eta_{ij} =
\alpha_j$, the Poisson MLE is $\hat\alpha_j = \log\sum_i c_{ij} -\break 
\log
\sum_i e^{\hat\mu_i}{=} \log( \sum_i c_{ij}/M )$ where $M =
\sum_i
m_i$, and $g(\hat\mu_i) = m_i(\sum_j \sum_i c_{ij}/\break M -1) = 0$.
\item Consider a single
$v_i \in
\{0,1\}$ such that $\eta_{ij} = \alpha_j + v_i \varphi_j$. Write
$C_{vj} = \sum_{i: v_i=v} c_{ij}$ and $M_{v} = \sum_{i:
v_i=v} m_i = \sum_j C_{vj}$. Then the Poisson MLE are $\hat\alpha_j =
\log(C_{0j}/\break M_0)$ and $\hat\varphi_j = \log(C_{1j}/M_1) - \log
(C_{0j}/M_0)$,
so that $g(\hat\mu_i) = m_i (\sum_j C_{v_ij}/M_{v_i} -1 ) =0$.
\end{itemize}
Of course, these examples do not form a general result: the situation
is more
complicated with correlated covariates or under regularization. But they
illustrate analytically why we might expect the performance we have seen
empirically: \mbox{estimates} based upon $\hat\mu_i = \log m_i$ do not
suffer in
out-of-sample validation. The resulting benefit is huge, as using a plug-in
allows estimation of the Poisson regression equations to proceed in complete
isolation from each other. See the  \hyperref[MR]{Appendix} for an example
MapReduce [\citet{deanmapreduce2004}] implementation.

\subsection{\texorpdfstring{Parallel Poisson regressions.}{Parallel Poisson regressions}}
\label{GL}

Given baseline intensities fixed as $\hat\mu_i = \log m_i$, each of
our $d$
separate Poisson regressions has negative log likelihood proportional to
\begin{equation}
\label{obj} l(\alpha_j, \bolds{\varphi}_j) = \sum
_{i=1}^n \bigl[ m_i
e^{\alpha_j + \mathbf{v}_i'\bolds{\varphi_j}} - c_{ij}\bigl(\alpha
_j +
\mathbf{v}_i'\bolds{\varphi}_j\bigr) \bigr].
\end{equation}
You are free to use your favorite estimation technique for each parallel
regression. This section outlines our specific approach: ``gamma
lasso'' $L_1$ regularized deviance minimization.

In high-dimensional regression, it can be useful to regularize estimation
through a penalty on coefficient size. This helps to avoid over-fit and
stabilize estimation.
A very common form of regularization imposes $L_1$ coefficient costs
[i.e., the lasso of \citet{tibshiraniregression1996}], which, due to
a nondifferentiable cost spike at the origin, yields variable
selection: some
coefficient estimates will be set to exactly zero. Our results here use
\textit{weighted $L_1$ regularization}
\begin{equation}
\label{wl1} \hat\alpha_j,\hat{\bolds{\varphi}}_j =
\mathop{\operatorname{arg}\operatorname{min}}_{\alpha_j,\bolds{\varphi}_j}
\Biggl\{l(\alpha_j,\bolds{\varphi}_j)
+ n \lambda\sum_{k=1}^p
\omega_{jk} |\varphi_{jk} | \Biggr\} \qquad\mbox{where }
\lambda,\omega_{jk} \geq0.
\end{equation}
Penalty size $\lambda$ acts as a \textit{squelch} that determines
what you
measure as signal and what you discard as noise. In practice, since optimal
$\lambda$ is unknown, one solves a \textit{regularization path} of candidate
models minimizing (\ref{wl1}) along the grid $\lambda_1 >
\lambda_2 >\cdots> \lambda_T$. Inference is completed through selection
along this path, with optimal $\lambda_t$ chosen to minimize cross-validation
(CV) or information criteria (IC; e.g., Akaike's AIC) estimated
out-of-sample (OOS) deviance (i.e., to minimize the average error for a given
training algorithm when used to predict new data). Crucially, {\it
selection is
applied independently for each category $j$ regression}, so that only a single
set of coefficients need be communicated back to a head node.

Analysis in this article applies the \textit{gamma lasso} algorithm of
\citet{taddygamma2013}, wherein weights $\omega_j$ diminish as a
function of $|\hat\varphi_j|$.\footnote{The iteratively reweighted least
squares algorithm in Section~6 of \citet{taddygamma2013} applies
directly to Poisson
family regressions by setting each iteration's ``observation weights''
$e^{\eta_{ij}}$ and
``weighted response'' $\eta_{ij} + c_{ij}/e^{\eta_{ij}} - 1$.} In particular,
along the grid of~$\lambda_t$ squelch values,
\begin{equation}
\label{glweight} \omega^{t}_{jk} = \bigl(1 + \gamma\bigl|\hat
\varphi^{t-1}_{jk}\bigr| \bigr)^{-1} \qquad\mbox{for }
\gamma\geq0.
\end{equation}
This includes the standard lasso at $\gamma=0$. For $\gamma>0$ it provides
\textit{diminishing bias} regularization, such that strong signals are less
shrunk toward zero than weak signals. This yields
sparser $\hat{\bolds{\varphi}}$, which reduces storage and
communication needs, and can
lead to lower false discovery rates. In practice, a good default is
$\gamma=0$ (i.e., the lasso), but if that provides solutions that are
implausibly (or inconveniently) dense, one can experiment with
increasing $\gamma$.\footnote{All of our
results use the {\tt gamlr} implementation in {\tt R}. The glass-shard
example of
Section~\ref{FGL} sets $\gamma=0$ for direct comparison to a lasso penalized
alternative, while the Yelp fits of Section~\ref{YELP} all use $\gamma
=1$ for
more sparsity.}

For selection along the path, we
minimize a \textit{corrected AIC} [\citet{hurvichregression1989}]:
\begin{equation}
\mbox{AICc:}\quad-2l(\hat\alpha_j,\hat{\bolds{\varphi}}_j)
+ 2\mathit{df}_j\frac{n}{n-\mathit{df}_j-1},
\end{equation}
where $\mathit{df}_j$ is the estimated \textit{degrees of freedom} used to fit
$\{\hat\alpha_j,\hat{\bolds{\varphi}}_j\}$. This corrects the AIC's
tendency to
over-fit, and \citet{taddygamma2013} finds that AICc performs well
with the gamma lasso. In Section~\ref{FGL}, where computation costs
are very low, we also consider CV selection
rules: both CV1se, which chooses the largest $\lambda_t$ with mean OOS
deviance no more than one standard error away from minimum, and CVmin,
which chooses $\lambda_t$ at lowest mean OOS deviance.

See \citet{taddygamma2013} for much more detail on these techniques.
That article reviews
diminishing bias regularization by emphasizing its close relationship
to weighted $L_1$ penalization.

\section{\texorpdfstring{Example: Glass shards and a parallel softmax.}{Example: Glass shards and a parallel softmax}}
\label{FGL}

Our motivating big-$d$ applications have the characteristic that $m_i$ is
random, and usually pretty big. For example, text mining $m_i$ is the total
word count in document $i$, and web analysis $m_i$ would be the total
count of
sites visited by a browser. A Poisson model for $m_i$ is not farfetched.
However, we also find that DMR also does well in the more common
\textit{softmax classification} setting, where $m_i=1$ always. It thus
provides an
everyday speedup for classification tasks: even with small-$d$ response
categories, you'll be able to fit the model almost $d$ times faster in
distribution.\footnote{In shared-memory parallelization we observe
speedups close to linear in $d$, depending upon machine
architecture.} Thus, before moving to our Yelp case study, we look at the
surprisingly strong performance of DMR in a simple classification problem.

\begin{figure}[t]

\includegraphics{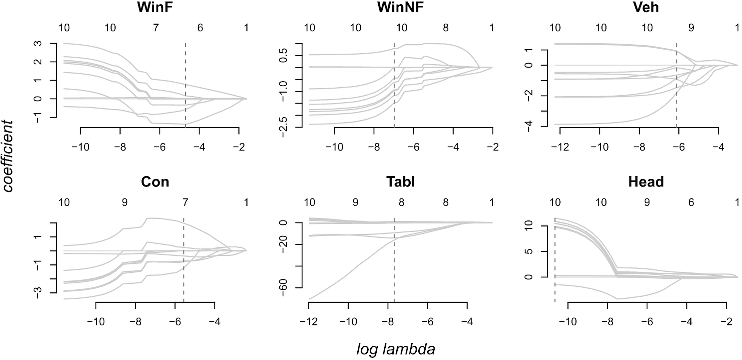}

\caption{\textit{Forensic glass}. Regularization paths
for each glass-type, with AICc selections marked.}\label{fglcoef}
\end{figure}

This example considers the small {\it forensic glass} data set from
\citet{venablesmodern2002}, available in the {\tt MASS} library
for {\tt R}
under the name {\tt fgl}.\footnote{For the code used in this example, type
{\tt help(dmr)} in {\tt R} after loading the {\tt distrom} library.} The data
are 214
observations on shards of glass. The response of interest is of 6 glass types:
window float glass ({\tt WinF}), window nonfloat glass ({\tt WinNF}), vehicle
window glass ({\tt Veh}), containers ({\tt Con}), tableware ({\tt
Tabl}), and
vehicle headlamps ({\tt Head}). Covariates for each shard are their refractive
index and \%-by-weight composition among 8 oxides. Figure~\ref{fglcoef}
shows Poisson regression regularization paths for each glass type,
with AICc selection marked by a vertical dashed line.
\begin{figure}[b]

\includegraphics{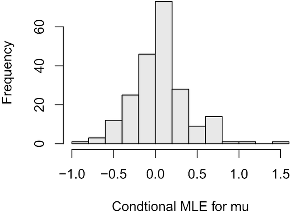}

\caption{\textit{Forensic glass}. The conditional MLEs $\mu^\star
_i$ implied at our DMR coefficient estimates.}
\label{fglmu}
\end{figure}

\begin{figure}

\includegraphics{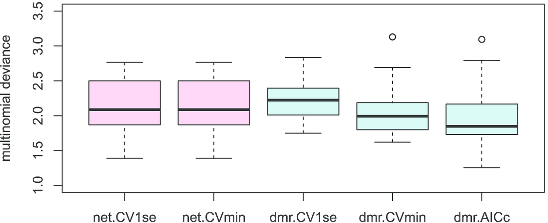}

\caption{\textit{Forensic glass}. OOS deviance samples in a
20-fold OOS experiment. The \textsf{net} models are
from {\tt glment} lasso multinomial logistic regression, and the
\textsf{dmr} models are our distributed multinomial
regression approximation. The applied penalty selection rule is
indicated after each model.}
\label{fglcv}
\end{figure}

The response here is a single category, such that $m_i = 1$ and
$\hat\mu_i=0$ for all $i$. This violates the assumption of Poisson
generation: $m_i=1$ is not random. For example, Figure~\ref{fglmu}
shows the conditional MLE $\mu_i^\star= \log(m_i/\sum_j e^{\hat
\alpha_j + \mathbf{v}_i'\hat{\bolds{\varphi}}_j} )$ at AICc selected
coefficients. The result is distributed around, but not equal to, the
assumed plug-in of $\hat\mu_i=0$ for all $i$. However, {\tt dmr} still
works: Figure~\ref{fglcv}
shows the distribution for OOS error in a 20-fold OOS experiment,
either using AICc or CV selection \textit{on each individual Poisson
regression}, against CV selected models from a lasso path for full
multinomial logistic (softmax) regression as implemented in the {\tt
glmnet} package for
{\tt R} [\citet{friedmanregularization2010}]. There are subtle
differences (e.g., AICc DMR selection has lower mean deviance with
higher variance), but the full multinomial fits ({\tt glmnet}) do not
have any clear advantage over the nearly $d$-times faster approximation
({\tt distrom}).

\section{\texorpdfstring{Yelp case study.}{Yelp case study}}
\label{YELP}

These data were supplied by the review site Yelp for a data mining
contest on
\url{kaggle.com}. The data are available at
\url{www.kaggle.com/c/yelp-recruiting/data}, and code for processing
and estimation is at
\url{github.com/TaddyLab/yelp}. We consider business, user, and review
data sets
in the yelp\_training\_data collection. The reviews, for all
sorts of
businesses, were recorded on January 19, 2013 for a sample of locations
near to
Phoenix, AZ. The goal of the competition was to predict the combined
number of
``funny,'' ``useful,'' or ``cool'' (f/u/c) votes that a given review
receives from
other users. Such information can be used by yelp to promote f/u/c reviews
before waiting for the users to grade them as such.

After detailing the data and model in Section~\ref{sec41}, we describe
a series of statistical analyses.
\begin{longlist}[Section~4.5:]
\item[Section~\ref{sec42}:] Investigate model fit under a range of
regularization
schemes, looking at how word loadings change with the relative weight of
penalty on variables of interest vs controls.
\item[Section~\ref{sec43}:] Use the ideas of ``sufficient reduction'' to
project text
through the model onto topics relevant to f/u/c votes or star ratings, and
interpret the resulting factor spaces.
\item[Section~\ref{sec44}:] Use the sufficient reductions in
prediction of the
number of f/u/c votes (i.e., the original {\tt kaggle} task), and compare
OOS performance against that of a word-count regression.
\item[Section~\ref{sec45}:] Use the sufficient reductions in
treatment effect
estimation---for the effect of user experience on rating---while controlling
for heterogeneity in review content.
\end{longlist}
By viewing text data as a big multinomial regression, we are able to address
all of the above (and resolve the effects of many collinear attributes on
review text) through a single model fit.

\subsection{\texorpdfstring{Data and model specification.}{Data and model specification}}\label{sec41}

The data are $n= 215{,}879$ reviews on 11{,}535 businesses by 43{,}873
users.\footnote{We have removed reviews with unknown user.} Review text
is split
on whitespace and tokenized into words (including combinations of punctuation:
potential emoticons). After stripping some common suffixes (e.g.,
``s,'' ``ing,''
``ly'') and removing a very small set of stopwords (e.g., ``the,''
``and,'' ``or''),
we count frequencies for $d= 13{,}938$ words occurring in more than 20
(${<}0.01\%$) of the reviews (total word count is $M= 17{,}581{,}214$).
Metadata
includes review, business, and user attributes:
\begin{itemize}
\item{\tt stars}: review star rating (out of 5), from which we subtract
the business average rating.
\item Review counts for {\tt funny}, {\tt useful}, or {\tt cool} votes. We
divide these by the square root of review age, which yields metrics
roughly uncorrelated with the posting date.
\item{\tt usr.count}: a user's total number of reviews at time of
posting the given review.
\item{\tt usr.stars}: a user's average star rating across all of their reviews.
\item A user's average {\tt usr.funny}, {\tt usr.useful}, or {\tt
usr.cool} votes
per review.
\item Business average star rating {\tt biz.stars} and review count
{\tt biz.count}.
\item Business location among 61 possible cities surrounding (and
including) Phoenix\hspace*{-0.7pt}.
\item Business classification according to Yelp's nonexclusive (and
partially user generated) taxonomy. We track membership for
333 categories containing more than 5 businesses.
\end{itemize}
This yields 405 variables for each review. We also specify random
effects\footnote{We call these random, rather than fixed, effects
because they
are estimated under a penalty which shrinks them toward an overall
mean. They will be estimated and we do not marginalize over them.}
for
each of the 11{,}535 businesses, leading to total attribute
dimension $p= 11{,}940$.
Data components are the $n \times d$ document--term
matrix $\mathbf{C}$, the $n$-vector of row-totals $\mathbf{m}$, and
the $n\times
p$ attribute matrix $\mathbf{V}$.

We split each row of the attribute matrix into two elements: $\mathbf
{a}_i$, the
11 numeric review attributes from {\tt stars} through
{\tt biz.count}, and $\mathbf{b}_i$, a length-11{,}929 vector of dummy
indicators for business identity,
location, and Yelp classification. This is done to differentiate the
variables we deem of primary interest ($\mathbf{a}_i$) from those
which we
include as controls ($\mathbf{b}_i$); write $\mathbf{V} = [\mathbf
{A} \ \ \mathbf{B}]$
as the
resulting partition. Columns of $\mathbf{A}$ are normalized to have mean
zero and
variance one. The multinomial regression of (\ref{bigmn}) is adapted by
similarly splitting each $\bolds{\varphi}_j = [ {\bolds{\varphi}_j^a},
{\bolds{\varphi}_j^b}]$ and rewriting category intensities
$
\eta_{ij} = \alpha_j + \mathbf{a}_i'\bolds{\varphi}^a_j + \mathbf
{b}_i'\bolds{\varphi}^b_j$.

\subsection{\texorpdfstring{Multinomial model fit and interpretation.}{Multinomial model fit and interpretation}}\label{sec42}

Following the recipe of Section~\ref{GL},
each word's Poisson regression is estimated
\begin{equation}
\label{yelpobj} \hat\alpha_j,\hat{\bolds{\varphi}}_j =
\mathop{\operatorname{arg}\operatorname{min}}_{\alpha_j,\bolds
{\varphi}_j} \biggl\{l(\alpha_j,\bolds{\varphi}_j) + n \lambda
\biggl[\sum_k
\omega^a_{jk} \bigl|\varphi^a_{jk} \bigr| +
\frac{1}{\tau
}\sum_k \omega^b_{jk}
\bigl|\varphi^b_{jk} \bigr| \biggr] \biggr\},
\end{equation}
where\vspace*{1pt} $l(\alpha_j, \bolds{\varphi}_j) = \sum_{i=1}^n [ m_i
e^{\alpha
_j +
\mathbf{a}_i'\bolds{\varphi}_j^a+ \mathbf{b}_i'\bolds{\varphi
}_j^b} - c_{ij}(\alpha_j +
\mathbf{a}_i'\bolds{\varphi}_j^a+ \mathbf{b}_i'\bolds{\varphi
}_j^b) ]$.
The \textit{relative penalty weight} $\tau> 0$ controls differential
regularization between the target variables and the controls. At larger
$\tau$
values, there is less penalty on $\bolds{\varphi}_j^b$ and the effect of
$\mathbf{b}_i$ on $c_{ij}$ has less opportunity to pollute our
estimate for
$\bolds{\varphi}_j^a$. That is, $\hat{\bolds{\varphi}}_j^a$ becomes
more purely
a {\it
partial effect}. At the extreme of $\tau= \infty$, any collinearity with
$\mathbf{b}_i$ is to be completely removed from the estimated $\hat{\bolds{\varphi}}_j^a$.

As outlined in  the \hyperref[MR]{Appendix}, counts for the 14k words are
partitioned
into 256 files. Each file is then read by one of 64 workstations, which
itself uses 16 cores in parallel to run through the Poisson
regressions. Each
individual regression is a full gamma lasso path solution over grids of 100
$\lambda_t$ squelch values, with weights $\omega_{jk}^{at},\omega_{jk}^{bt}$
updated as in (\ref{glweight}) under $\gamma=1$, and AICc selected
coefficients are then written to file. The entire problem (including the
sufficient reduction projection of our next section) takes around $1/2$ hour.

\begin{figure}

\includegraphics{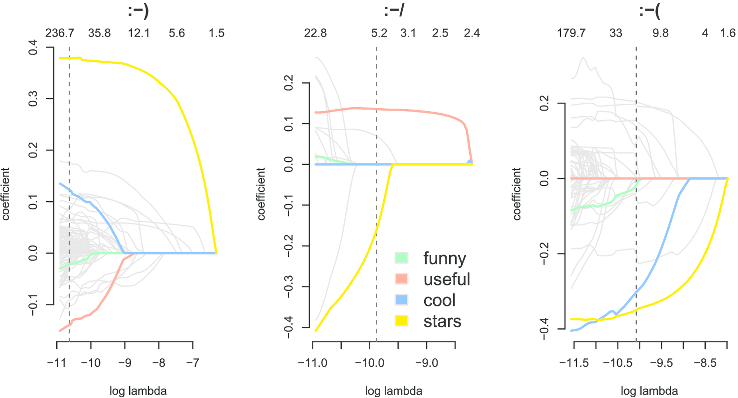}

\caption{\textit{Yelp}. Poisson regression regularization paths for
counts of the tokens {\tt:$\!$-$\!$)},
{\tt:$\!$-$\!$/}, and
{\tt:$\!$-$\!$(} under relative penalty weight $\tau=2$. Coefficient values
have been
multiplied by the corresponding covariate standard deviation. The legend
highlights select covariates, regression degrees of freedom are on the top
axis, and our AICc selected estimates are marked with vertical dashed
lines.}
\label{yelpir}
\end{figure}

Regularization paths for a few of the Poisson regressions, estimated under
$\tau=2$ relative penalty weight, are shown in Figure~\ref{yelpir}.
Coefficient values are scaled to the effect of 1sd change in
the corresponding attribute. We see, for example, that at our AICc selection
the effect of a 1sd increase in review stars multiplies the expected
count (or
odds, in the multinomial model) for the happy face
\texttt{:$\!$-$\!$)} by around $\exp0.38 \approx1.46$, the ``hmmm'' face {\tt
:$\!$-$\!$/} by $\exp
-0.15 \approx0.86$, and the sad face {\tt:$\!$-$\!$(} by $\exp-0.35
\approx0.7$.
Notice that {\tt:$\!$-$\!$/} and {\tt:$\!$-$\!$(} both occur more often in low-star
(negative) reviews, but that {\tt:$\!$-$\!$/} is associated with useful
content, while
{\tt:$\!$-$\!$(} is uncool.

The relative penalty divisor $\tau$ allows us to specify the amount of word
count variation that is allocated the control variables in $\mathbf
{B}$. Such
differential penalization is a powerful tool in Big Data analysis, as it
allows one to isolate partial effects in messy overdetermined systems.
Unfortunately, unlike for $\lambda$, we have no objective criterion
with which to
choose $\tau$. Since it weights the penalty on variables whose effect we
would like to \textit{remove} from our targeted coefficients, one could
argue that
$\tau= \infty$ is the optimal choice. In practice, however, this can
lead to
inference for the coefficients of interest that is dependent upon only
a small
subset of documents (since variation in the others is saturated by the
controls). We advise using whatever prior knowledge is available to
evaluate the appropriateness of results under a range of $\tau$.

For example, Table~\ref{topwords} investigates fit under increasing
$\tau$.
The numbers of nonzero $\hat\varphi^a_{jk}$ (i.e., deemed useful for OOS
prediction by AICc) are decreasing with $\tau$ for all attributes.
This is
because $\mathbf{B}$ accounts for more variation in $\mathbf{C}$ at
higher $\tau
$, and
there is little residual variation left for $\mathbf{A}$. Here, $\tau
=2$ yields
top words only indirectly associated with our attributes (e.g., \textit
{prik} is
positive because Thai food gets high ratings), while full $\tau=\infty$
control leads to near perfect fit and infinite likelihoods conditional on
$\mathbf{B}$ alone. To our eye, $\tau=20$ manages a good balance:
there remain
many significant $\hat\varphi^a_{jk}\neq0$, but the model has
avoided loading
words that are not directly associated with the given attributes. This
fit is
used in the remainder of our study.

\begin{table}
\tabcolsep=0pt
\caption{Top 10 words by loading on review characteristics, as a
function of relative
penalty weight $\tau$. The top row for each attribute corresponds to
terms ordered by
marginal correlations}\label{topwords}
\begin{tabular*}{\tablewidth}{@{\extracolsep{\fill}}lccl@{}}
\hline
& $\bolds{\tau}$ & $\bolds{\hat\varphi\neq0}$
& \textbf{Top ten words by loading}\\
\hline
& \multicolumn{2}{c}{Marginal} & \textit{Great love amaz
favorite deliciou best awesome alway perfect excellent} \\
$+$Stars & \phantom{00}2 & 8440 & \textit{Unmatch salute :-)) prik
laurie pheonix trove banoffee exquisite sublime} \\
& \phantom{0}20 & 3077 & \textit{Heaven perfection gem divine amaz die
superb phenomenal fantastic}\\
&&& \textit{deliciousnes} \\
& 200 & \phantom{0}508 & \textit{Gem heaven awesome wonderful amaz
fantastic favorite love notch fabulou} \\[6pt]
& \multicolumn{2}{c}{Marginal} & \textit{Not worst ask
horrib minut rude said told would didn} \\
$-$Stars & \phantom{00}2 & 8440 & \textit{Rude livid disrespect
disgrace inexcusab grossest incompet audacity unmelt}\\
&&& \textit{acknowledge} \\
& \phantom{0}20 & 3077 & \textit{Rude incompet unaccept unprofession
inedib worst apolog disrespect insult}\\
&&& \textit{acknowledge} \\
& 200 & \phantom{0}508 & \textit{Worst horrib awful rude inedib terrib
worse tasteles disgust waste} \\[6pt]
& \multicolumn{2}{c}{Marginal} & \textit{You that know like
your yelp $\ldots$ what don who} \\
Funny & \phantom{00}2 & 6508 & \textit{Dimsum rue reggae acne
meathead roid bong crotch peni fart} \\
& \phantom{0}20 & 1785 & \textit{Bitch shit god dude boob idiot fuck
hell drunk laugh} \\
& 200 & \phantom{0}120 & \textit{Bitch dear god hell face shit hipst
dude man kidd} \\[6pt]
& \multicolumn{2}{c}{Marginal} & \textit{That yelp you thi
know biz-photo like all http ://} \\
Useful & \phantom{00}2 & 5230 & \textit{Fiancee rife dimsum maitre
jpg poultry harissa bureau redirect breakdown} \\
& \phantom{0}20 & \phantom{0}884 & \textit{biz-photo meow harissa
www bookmark :-/
http :// (?), tip} \\
& 200 & \phantom{00}33 & \textit{www http :// com factor already final
immediate ask hope} \\[6pt]
& \multicolumn{2}{c}{Marginal} & \textit{Yelp you that
biz-photo http :// www know like your} \\
Cool & \phantom{00}2 & 4031 & \textit{Boulder lewi rogue lagunita
wanton celebratory hanker politic mozzerella}\\
&&& \textit{onsite} \\
& \phantom{0}20 & \phantom{0}577 & \textit{Userid htm cen rand poem
sultry arlin
brimm cubic inspiration} \\
& 200 & \phantom{00}11 & \textit{Biz-photo select yelp along certain fil
chose house} \\
\hline
\end{tabular*}
\end{table}

\subsection{\texorpdfstring{Sufficient reduction.}{Sufficient reduction}}\label{sec43}

The previous section's coefficient estimates, resolving a complex
system of
relationships between words and attributes, provide a rich basis for
storytelling and exploratory analysis. For many, this is either the end goal
or a jumping-off point (e.g., to experiments testing hypotheses generated
in exploration). But in our practice, a primary reason for fitting big
multinomial models is as a tool for \textit{dimension reduction},
mapping from
the original $d$-dimensional text down to univariate indices that
contain all
information relevant to a given attribute.

\citet{cookfisher2007} outlines use of regression models with
high-dimensional response as a map to project from that response onto
interesting covariates. \citet{taddymultinomial2013} extends the
idea in
our context of big multinomials, motivated by applications in text analysis.
Both of these articles are focused on {\it inverse regression} (IR), a
technique wherein the fitted model map is applied for prediction of unobserved
covariates (e.g., the votes associated with new review text, as in
Section~\ref{sec44}). However, the IR algorithms are prefaced on a
more basic
concept of {\it sufficient reduction} (SR), which is useful beyond its
application in IR prediction.

Consider observation $\mathbf{c}_i$ from a $d$-dimensional exponential family
linear model, with natural parameter $\bolds{\eta}_i = [\eta_{i1}
\cdots
\eta_d]'$, $\eta_{ij} = \alpha_j + \mathbf{v}_i\bolds{\varphi
}_j$, such that
\begin{equation}
\label{nef} \mathrm{p}(\mathbf{c}_i) = h(\mathbf{c}_i)\exp
\bigl[\mathbf{c}_i'\bolds{\eta}_i + A(\bolds{\eta}_i) \bigr],
\end{equation}
where $h$ is a function of only data (not $\bolds{\eta}_i$) while $A$
is a
function of only parameters (not $\mathbf{c}_i)$. Both the full
multinomial logistic regression model (conditional upon $m_i$) or our
independent Poisson's model (conditional upon $\hat\mu_i$) can be
written as in~(\ref{nef}). Then with $\bolds{\Phi} = [\bolds{\varphi}_1 \cdots
\bolds{\varphi}_d]$ the $p\times d$ matrix of regression
coefficients, we get
\begin{equation}
\label{srproof} \mathrm{p}(\mathbf{c}_i) = h(\mathbf
{c}_i)e^{\mathbf{c}_i'\bolds{\alpha}}
\exp\bigl[\mathbf{c}_i'\bolds{\Phi}'
\mathbf{v}_i + A\bigl(\bolds{\Phi}'\mathbf{v}_i\bigr)
\bigr] = \tilde h(\mathbf{c}_i)g(\bolds{\Phi}\mathbf{c}_i,
\mathbf{v}_i),
\end{equation}
so that the likelihood factorizes into a function of $\mathbf{c}_i$
only and
another function of $\mathbf{v}_i$ that depends upon $\mathbf{c}_i$ only
through the
projection $\bolds{\Phi}\mathbf{c}_i$. This implies that,
conditional upon
the regression parameters, $\bolds{\Phi}\mathbf{c}_i$ is a {\it
sufficient statistic} for $\mathbf{v}_i$. That is, $\mathbf{v}_i
\perp\!\!\!\perp\mathbf{c}_i \vert
\bolds{\Phi}\mathbf{c}_i$.

We call $\mathbf{z}_i = \bolds{\Phi}\mathbf{c}_i$ an SR {\it
projection}. In
practice, we
work with estimated SR projections $\mathbf{z}_i = \hat{\bolds{\Phi}}\mathbf{c}_i$ and
hope that $\hat{\bolds{\Phi}}$ has been estimated well\vspace*{1pt} enough for
$\mathbf{z}_i$
to be
a useful summary [see \citet{taddyrejoinder2013} for discussion].
In that
case, $\hat{\bolds{\Phi}}$ provides a linear map from text into the
$p$-dimensional
attribute space. This works just like the {\it rotation} matrix from common
principal components analysis except that, instead of mapping into latent
factors, $\bolds{\Phi}$ projects into observed attributes. The resulting
$\mathbf{z}_i$ are model-based sufficient statistics, useful in the
same roles
as a traditional sufficient statistic (like $\bar x$). For example, to predict
$v_{ik}$ from $\mathbf{c}_i$ we can work with univariate $z_{ik}$ instead
of the
$d$-dimensional original text. In general, SR projections are a simple
way to
organize information in Big Data systems. When new text $\mathbf{c}_i$ arrives,
one need just feed it through $\bolds{\Phi}$ to obtain $\mathbf
{z}_i$ indices which
can be summarized, plotted, and analyzed as desired.

It is important to emphasize that, since estimated loadings
$\hat\varphi_{ik}$ are partial effects (influence of other attributes has
been controlled for), $z_{ik}$ will also correspond to partial rather
than marginal association. As another way to see this, note that the
factorization in (\ref{srproof}) is easily manipulated to show
sufficiency for each individual $z_{ik}$ conditional on $\mathbf
{v}_{i,-k}$, our
vector of attributes omitting the $k$th. Thus, SR reduces
dimension into a space of information \textit{directly} relevant to an
attribute
of interest, where influence of text variation due to other
attributes has been removed or minimized. Consider the correlation
matrices in
Figure~\ref{zcor}. The original vote attributes are highly positively
correlated,
while the text projections are either nearly independent (e.g., {\tt useful}
against either {\tt funny} or
{\tt cool}) or strongly negatively correlated ({\tt funny} and {\tt
cool}). This
suggests that there are underlying factors that encourage \textit
{votes in any
category}; only after controlling for these confounding factors do we
see the
true association between f/u/c content. Similarly, all vote attributes are
uncorrelated with star rating, but for the text projections we see both negative
({\tt funny}, {\tt useful}) and positive ({\tt cool}) association.

\begin{figure}
\footnotesize{\textbf{Correlation matrices}}\vspace*{3pt}
\begin{center}
\footnotesize{
\begin{tabular}{@{}r|cccc@{}r|cccc@{}}
\multicolumn{4}{l}{\textit{Attributes} ($\mathbf{v}$)} & &
\multicolumn{4}{l}{\textit{\hskip2cm Text projections} ($\mathbf
{z}$)} \\
\multicolumn{1}{c}{}& f & u &c & $\star$ & \multicolumn
{1}{c}{}&f\vspace*{-3pt} & u
&c & $\star$\\[2pt]
\cline{2-5} \cline{7-10}
f & 1\phantom{.0} & 0.7 & 0.8 & 0 & \hskip2cm f & 1 & $-$0.1 & $-$0.7 & $-$0.4\\
u & 0.7 & 1\phantom{.0} & 0.9 & 0 & \hskip2cm u & $-$0.1 &
1 & \phantom{$-$}0.1 & $-$0.2\\
c & 0.8 & 0.9 & 1\phantom{.0} & 0 & \hskip2cm c & $-$0.7 & \phantom
{$-$}0.1 & 1 & \phantom{$-$}0.5\\
$\star$ & 0\phantom{.0} & 0\phantom{.0} & 0\phantom{.0} & 1 &
\hskip2cm $\star$ & $-$0.4 & $-$0.2 & \phantom{$-$}0.5 &
1\\
\end{tabular}
}
\end{center}
\caption{Correlation for the original review attributes
in $\mathbf{v}$ (left) and for $\mathbf{z}$ (right) SR text
projection. Here
$f$ denotes either the observed ($v_{\tt funny}$) or sufficient
reduction for ($z_{\tt funny}$) the number of funny votes per square
root review age, similarly with $u$ for useful votes and $c$ for cool
votes, and $\star$~denotes the observed and SR for the number of review
stars.}
\label{zcor}
\end{figure}

The three 50--100 word reviews in Figure~\ref{threerev} provide further
illustration. A single review (bottom) of a historical site scores
highest in
${\tt funny}$ and ${\tt useful}$ attributes (and also in ${\tt cool}$). The
review is neither dry nor useless, but we imagine its high vote count
has been
influenced by other factors, for example, the page is heavily viewed or
people who
read reviews of national parks are more likely to vote. In contrast,
the two
reviews identified through SR projections as having the most funny or useful
content appear to us as more directly related to these concepts. The funny
review, for a pizza restaurant, is a fictional comedic story. The useful
review contains a high proportion of business photos (\textit
{biz-photo}), which
the multinomial model has identified as directly useful. The machine-learned
text projections are able to detect humor and helpfulness distinct from
the other
factors that lead to increased user votes.

\begin{figure}
\begin{center}\small
\framebox[0.95\textwidth]{
\parbox{0.9\textwidth}{
\vskip.5cm
{\bf Funniest 50--100 word review, by SR projection ${\mathbf{z}}_{\tt
funny}$.}

\vskip.2cm
{\it Dear La Piazza al Forno: We need to talk. I do not quite know how
to say this so I'm just going to come out with it. I've been seeing
someone else. How long? About a year now. Am I in love? Yes. Was it
you? It was. The day you decided to remove hoagies from your lunch
menu, about a year ago, I'm sorry, but it really was you$\ldots$ and
not me.
Hey$\ldots$ wait$\ldots$ put down that pizza peel$\ldots$ try to
stay calm$\ldots$ please?
[Olive oil container whizzing past head] Please! Stop throwing shit at
me$\ldots$ everyone breaks up on social media these days$\ldots$ or
have not you
heard? Wow, what a Bitch!}

\vskip.5cm
{\bf Most useful 50--100 word review, by SR projection ${\mathbf
{z}}_{\tt useful}$.}

\vskip.2cm
{\it We found Sprouts shortly after moving to town. There's a nice
selection of Groceries \& Vitamins. It's like a cheaper, smaller
version of Whole Foods.
[biz-photo] [biz-photo] We shop here at least once a week. I like their
selection of Peppers$\ldots.$ I like my spicy food!
[biz-photo][biz-photo][biz-photo] Their freshly made Pizza is not too
bad either. [biz-photo] Overall, it's a nice shopping experience for
all of us. Return Factor---100\%.}

\vskip.5cm
{\bf Funniest and most useful 50--100 word review, as voted by Yelp
users\\ (votes normalized by square root of review age).}

\vskip.2cm{\it
I use to come down to Coolidge quite a bit and one of the cool things I
use to do was come over here and visit the ruins. A great piece of
Arizona history! Do you remember the Five C's? Well, this is cotton
country. The Park Rangers will tell you they do not really know how old
the ruins are, but most guess at around 600 years plus. But thanks to a
forward thinking US Government, the ruins are now protected by a 70
foot high shelter. Trust me, it comes in handy in July and August, the
two months I seem to visit here most. LOL. I would also recommend a
visit to the bookstore. It stocks a variety of First Nation history, as
well as info on the area.
\url{http://www.nps.gov/cagr/index.htm}. While you are in Coolidge, I would
recommend the Gallopin' Goose for drinks or bar food, and Tag's for
dinner. Both are great!}
\vskip.5cm
}}
\end{center}
\caption{Illustration of the information contained in
sufficient projections
$\mathbf{z}$. The top two reviews are those, among all where $m\in(50,100)$,
with highest SR projection scores into the {\tt funny} and {\tt useful}
attribute spaces. For comparison, we also show the single 50--100 word review
with highest values for both $v_{\tt funny}$ and $v_{\tt useful}$
(recall that
these are vote totals per square root review age). Note that, since variance
of $\mathbf{z}$ increases with $m$, high scoring reviews tend to be
longer. One
can also, as in
\citeauthor{taddymultinomial2013} \textup{(\citeyear{taddymultinomial2013})}, divide the SR projections by document
length and work with normalized $z/m$.
On this scale, the funniest review is ``\textit{Holy Mother of God}'' and
the most
useful review is ``\textit{Ask for Nick}!''}
\label{threerev}
\end{figure}

\subsection{\texorpdfstring{Inverse regression for prediction.}{Inverse regression for prediction}}\label{sec44}

Multinomial-based SR projections were originally motivated by \citet{taddymultinomial2013} for their use in {\it multinomial inverse
regression} [MNIR; see also \citet{taddymeasuring2013}]. Say $v_{iy}$,
some element of the attribute vector $\mathbf{v}_i$, is viewed as a
``response'' to be predicted for future realizations. For example, in
the original {\tt kaggle} Yelp contest the goal was to predict
$v_{i,\tt funny}$, $v_{i,\tt useful}$, or $v_{i,\tt cool}$---the vote
attributes. In such applications, an MNIR routine would use the SR
projection into $v_{iy}$, $z_{iy} = \sum_j \hat\varphi_{jy} c_{ij}$,
to build a {\it forward regression} that predicts $v_{iy}$ from
$z_{iy}$, $\mathbf{v}_{i,-y}$ (attributes omitting $y$), and
$m_i$.\footnote
{The SR result that applies here is $v_{iy} \perp\!\!\!\perp\mathbf
{c}_i \vert
z_{iy}, \mathbf{v}_{i,-y}, m_i$. Since sufficiency for $z_{iy}$ from the
multinomial factorization is \textit{conditional upon $m_i$}, these
document totals need to be conditioned upon in forward regression.}
This $p+1$ dimensional regression replaces the $d+p-1$ dimensional one
that would have been necessary to predict $v_{iy}$ from $\mathbf{v}_{i,-y}$
and $\mathbf{c}_{i}$, the original text counts.

Estimating an \textit{inverse} regression in order to get at another
\textit{forward} regression may seem a strange use of resources. But there
are a
variety of reasons to consider MNIR. Computationally, through either the
techniques of this article or the collapsing of
\citet{taddymultinomial2013}, the multinomial regression estimation
can occur
in distribution on many independent machines. This is useful when the full
count matrix $\mathbf{C}$ is too big to fit in memory.
Another reason to use MNIR is for statistical efficiency when $d$ is big
relative to $n$. Assuming a multinomial distribution for $\mathbf{c}_i
\vert
\mathbf{v}_i$ introduces information into the estimation problem (a
less generous
term is ``bias''). In particular, it implies that each of the $M
= \sum_i m_i$ counts are independent observations, such that the
sample size
for learning $\bolds{\Phi}$ becomes $M$ rather than $n$. That is,
estimation variance decreases with the number of words rather than the number
of documents [see \citet{taddyrejoinder2013}].

\begin{table}
\caption{\textit{Yelp}. Out-of-sample $R^2$ in
prediction for vote attributes (normalized by root review age) in
5-fold~CV. The top row shows a standard lasso regression from the vote
attribute onto text and all nonvote attributes, while the bottom row
holds results for MNIR followed by lasso regression from the vote
attribute onto review length ($m_i$), nonvote attributes, and the
corresponding univariate~SR~projection}\label{yelpcv}
\begin{tabular*}{\tablewidth}{@{\extracolsep{\fill}}lccccc@{}}
\hline
&\multicolumn{2}{c}{\textbf{Forward regression}}&
\multicolumn{3}{c}{\textbf{Average out-of-sample} $\bolds{R^2}$}\\[-4pt]
& \multicolumn{2}{l}{\hrulefill} & \multicolumn{3}{l@{}}{\hrulefill
}\\
& \textbf{Input variables} & \textbf{Dimension} & \textbf{\texttt{funny}} & \textbf{\texttt{useful}} & \textbf{\texttt{cool}}\\
\hline
Standard lasso & Nonvote attributes, $\mathbf{C}$ & 25{,}876 & 0.308 & 0.291
& 0.339\\
MNIR${}+{}$lasso & Nonvote attributes, $\mathbf{z}$, $\mathbf{m}$ &
11{,}940 & 0.316
& 0.296 & 0.341\\
\hline
\end{tabular*}
\end{table}

As an illustration, Table~\ref{yelpcv} shows results for prediction of
individual f/u/c vote attributes, both through MNIR with lasso forward
regression
and for a standard lasso onto the full text counts. That is, MNIR fits
$\mathbb{E}[v_{iy}] = \beta_0 +
[\mathbf{v}_{i,-{\mathrm{f/u/c}}},m_i,z_{iy}]'\bolds{\beta}$ while the
comparator fits
$\mathbb{E}[v_{iy}] = \beta_0 + [\mathbf{v}_{i,-{\mathrm{f/u/c}}},\mathbf{c}_i]'\bolds{\beta}$,
where $\mathbf{v}_{i,-{\mathrm{f/u/c}}}$ denotes all nonvote
attributes. For MNIR
each $\hat{\bolds{\Phi}}$ (hence, $z_{iy}$) is also estimated using
only the
training sample, and in both cases prediction rules were selected via AICc
minimization along the $L_1$ regularization path. We see that
MNIR forward regression, replacing 13{,}938 covariates from $\mathbf{c}_i$
with just
the two numbers $z_{yi}$ and $m_i$, does not suffer against the full lasso
comparator (indeed, it is very slightly better in each case). Such performance
is typical of what we have observed in application.\footnote{In
\citet{taddymultinomial2013}, the MNIR routines more
significantly outperform
lasso comparators in OOS prediction. However, the data sets used in
that paper
are both very small, with $M\gg n$. Thus, our statistical efficiency
argument---that for MNIR estimation variance decreases with $M$
instead of $n$---is
working heavily in favor of MNIR. Here, even though $M > n$, vocabulary
size $d$ is smaller than $n$ and linear regression is already plenty
efficient.} This is not evidence that the text counts do not matter: each
full lasso estimates at least 4000 terms having nonzero coefficients.
Rather, the multinomial model is a good enough fit that the factorization
results of (\ref{srproof}) apply and all relevant information is contained
in the SR projection.\footnote{We have also found success applying
nonlinear learning (e.g., trees) in forward regression after SR projection.
Methods that are too expensive or unstable on the full text work
nicely on
the reduced dimension subspace.}

Note that the MNIR forward regression ignores projection from text onto any
other nonvote attributes. This is because those attributes are conditioned
upon in forward regression. Indeed,
\citeauthor{taddymultinomial2013} (\citeyear{taddymultinomial2013,taddyrejoinder2013}) argue that, in prediction
for a single variable, you only need fit the multinomial dependence between
counts and that single variable. This yields SR projection based on marginal
association, which can work as well as that based on partial
association for simple predictions. The benefit of fitting models for
high-dimensional $\mathbf{v}_i$ is that we are then able to interpret
the resulting
partial effects and SR projections, as in Sections~\ref{sec42}--\ref
{sec43}. It is also useful in more structured prediction
settings, as in the next section.

\subsection{\texorpdfstring{Confounder adjustment in treatment effect
estimation.}{Confounder adjustment in treatment effect
estimation}}\label{sec45}

In our final application, we illustrate use of SR projections as
convenient low-dimensional {\it controls} in treatment effect
estimation. The task here has
a particular attribute, say $t$, whose effect on another, say $y$, you
want to
estimate. You want to know what will happen to $y$ if $t$ changes
independently from the other attributes. Unfortunately, everything is
collinear in the data and both $y$ and $t$ could be correlated to other
unobserved confounders. Your best option is to estimate the treatment
effect---that of $t$ on~$y$---while controlling for observable potential
confounders. In text analysis, this includes controlling for the text content
itself.

Consider estimating the effect of a user's experience---the number of reviews
that they have written---on their expected rating. That is, are experienced
users more critical, perhaps because they have  become more discerning? Or do
they tend to give more positive reviews, perhaps because community norms
encourage a high average rating? It is hard to imagine getting firm evidence
in either direction without running a randomized trial---we will
always be
worried about the effect of an omitted confounder. However, we can try our
best and condition on available information. In particular, we can
condition on content to ask the question: even given the same review
message, would an experienced user give more or less stars than a newbie?

The response attribute, $v_{iy}$, is \textit{star rating}. The treatment,
$v_{it}$, is the log {\it number of reviews} by the author
(including the current review, so never less than one). Results for
estimation of the effect of $v_{it}$ on $v_{iy}$, conditioning on different
control variables, are detailed in Table~\ref{treats}. A na\"ive estimate
for the effect of experience on rating, estimated through the marginal
regression $\mathbb{E}[v_{iy}] = \beta_0 + v_{it}\gamma$, is a $\hat
\gamma=
0.003$ increase in number of stars per extra unit log review count. Use
$\mathbf{v}_{i,-yt}$ to denote all other attributes. Then an improved
estimate of
the treatment effect is obtained by fitting $\mathbb{E}[v_{iy}] =
\beta_0 +
v_{it}\gamma+
\mathbf{v}_{i,-yt}'\bolds{\beta}$, which yields the much larger
$\hat\gamma= 0.015$.

Finally, we would like to control for $\mathbf{c}_i$, the review content
summarized as word counts. It would also be nice to control for content
interacting with attributes since, for example, positive content for a
restaurant might imply a different star rating boost than it does for a
bowling alley.
Unfortunately, interacting 13{,}938 dimensional $\mathbf{c}_i$ with
the 333
business categories yields almost 4.7 million regression coefficients.
This is more controls than we have observations. However, the SR
projections offer a low-dimensional alternative.
Write $z_{iy}$ and $z_{it}$ for the SR
projections onto response and treatment, respectively. Then sufficiency
factorization implies
\begin{equation}
v_{iy},v_{it} \perp\!\!\!\perp\mathbf{c}_i \vert
z_{iy},z_{it},m_i,\mathbf{v}_{i,-yt}.
\end{equation}
That is, the joint distribution of treatment and control is independent of
the text given SR projection into each. This suggests we can control for
review content, and its interaction with business classification,
simply by
adding to our conditioning set $[z_{iy},z_{it},m_i]$ and its
interaction with
business classification. The resulting regression, with around 13k control
coefficients instead of 4.7 million, yields the still larger treatment effect
estimate $\hat\gamma=0.02$.

\begin{table}
\tabcolsep=0pt
\caption{Estimated effect ``$\gamma$'' of user
experience (log
number of reviews) on number of stars rated. Each corresponds to different
levels of confounder adjustment. The effects are all AICc selected estimates
along a $\gamma=10$ (very near to $L_0$) gamma lasso regularization path,
where \textit{all of the other regression coefficients} were
unpenalized. Thus,
they are significant, in the sense that the AICc deems $v_{it}$ useful for
predicting $v_{iy}$ even after all variation explained by confounders
has been
removed}\label{treats}
\begin{tabular*}{\tablewidth}{@{\extracolsep{\fill}}lccc@{}}
\hline
& \textbf{Marginal} & \textbf{Conditional on attributes only} &
\textbf{Adding and interacting
text SR} \\
\hline
Effect estimate & 0.003 & 0.015 & 0.020
\\
\hline
\end{tabular*}
\end{table}

\section{\texorpdfstring{Discussion.}{Discussion}}
\label{END}

Distributed estimation for multinomial regression allows such models to be
applied on a new scale, one that is limited only by the number of machines
you are able to procure. This is an important advance not only for our
motivating text
analysis applications, but also for any other setting of high-dimensional
multinomial modeling. This includes any softmax classification model.

One message of this paper has been that Poisson factorization enables fast
estimation of multinomial distributions. It has been pointed out to us that,
in unstructured data analysis,\vadjust{\goodbreak} a Poisson seems little more arbitrary
than a
multinomial model. Equation (\ref{embed}) clarifies this issue: the only
additional assumption one makes by working with independent Poissons is that
the aggregate total, $m_i$, is Poisson. We have attempted to mitigate the
influence of this assumption, but that is unnecessary if you consider the
Poisson a fine model in and of itself.

Finally, we wish to emphasize the relative simplicity of this approach.
Although this article describes models for complex language systems,
and it
may not seem to the reader that we are providing anything ``simple,''
almost all
of this material is \textit{just Poisson regression.} We have used the
ideas of
\citet{taddygamma2013} and the gamma lasso to fit these regressions,
but any
generalized linear model estimator could have been applied. As
implemented in
this article, there are only two tuning parameters in the entire
system: the
gamma lasso weight $\gamma$, which can be safely fixed at zero (for lasso
regression) as a solid default; and the relative confounder-penalty divisor
$\tau$. Specification of $\tau$ is a clearly subjective
choice,\footnote{Except for when you have enough data to identify the
model with $\tau
= \infty$.}
but such subjectivity is inevitable in any structural inference that
does not
involve a random or pseudo-random experiment.

Too often, social scientists faced with text data will jump to latent space
models (e.g., topic models) as the first step in their analysis. Unless the
phenomena that they'd like to measure is a dominant source of variation in
word choice, these latent topics will be mostly irrelevant. The same
scientist faced with more familiar response variables---such as money
spent---would likely have used regression modeling with a mix
of observable covariates and fixed or random effects, instead of trying to
model any sort of latent space. Thus, without wanting to claim that
topic and
related models are not useful (they are very useful), we hope that this
article will give social scientists the option of using the same type of
regression tools for text analysis that they use successfully in their
nontext research.

\begin{appendix}\label{MR}
\section*{\texorpdfstring{Appendix: MapReduce details}{Appendix: MapReduce details}}

MapReduce (MR) is a recipe for analysis of massive data sets, designed to
work when the data itself is distributed: stored in many files on a
network of
distinct machines. The most common platform for MR is Hadoop paired
with a
distributed file-system (DFS) optimized for such operations (e.g.,
Hadoop DFS).

A MapReduce routine has three main steps: map, partition, and reduce. The
partition is handled by Hadoop, such that we need worry only
about map and reduce. The map operation parses unstructured data into a
special format. For us, in a text mining example, the mapper program will
take a document as input, parse the text into tokens (e.g., words), and output
lines of processed token counts: ``{\tt token document|count}.'' The pre-tab
item (our {\tt token}) is called a ``key.'' Hadoop's sort facility uses these
keys to send the output of your mappers to machines for the next step,
reducers, ensuring that all instances of the same key (e.g., the same word)
are grouped together at a single reducer. The reducer then executes some
operation that is independent-by-key, and the output is written to file
(usually one file per reducer).

DMR fits nicely in the MR framework. Our map step tokenizes your unstructured
data and organizes the output by token keys. Reduce then takes all
observations on a single token and runs a Poisson log regression,
applying the
gamma lasso with IC selection to obtain coefficient estimates. This recipe
is detailed in Algorithm \ref{mralgo}.

\begin{algorithm}[t]
\caption{MapReduce DMR}\label{mralgo}
\vskip.25cm
{\bf Map:} For each document, tokenize and count sums for each token.
Save the total counts $m_i$ along with attribute information $\mathbf
{v}_i$. Output {\tt token document|count}.

\vskip.25cm
Combine totals $m_i$ and attributes $\mathbf{v}_i$ into a single
table, say
{\tt VM}. This info can be generated during map or extracted in earlier
steps. Cache {\tt VM} so it is available to your reducers.

\vskip.25cm
{\bf Reduce:} For each token key ``$j$,'' obtain a regularization path
for Poisson regression of counts $c_{ij}$ on attributes $\mathbf{v}_i$ with
$\hat\mu_i = \log m_i$. Apply AICc to select a segment of coefficients
from this path, say $\hat{\bolds{\varphi}}_j$, and output nonzero elements
in sparse triplet format: {\tt word|attribute|phi}.

\vskip.25cm
Each reducer writes coefficients $\hat{\bolds{\varphi}}_j$ of interest to
file, and maintains a running total for SR projection, $\mathbf{z}_i
+\!\!=
\mathbf{c}_i'\hat{\bolds{\varphi}}_j$, output as say {\tt Z.r} for
the $r$th
reducer. When all machines are done we aggregate {\tt Z.r}
to get the complete projections.
\end{algorithm}

We have written this as a single MR algorithm, but other variations may
work better for your computing architecture. Our most common
implementation uses Hadoop to execute the map on a large number of
document files, but replaces the regression reduce step with a simple
write, to solid state storage ``midway'' at the University of Chicago's
Research Computing Center, of token counts tabulated by observation.
For example, given 64 reducer machines on AWS, the result is 64 text
tables on midway, with lines ``{\tt word|doc|count},'' each containing
\textit{all} nonzero counts for a subset of the vocabulary of tokens.
These files are small enough to fit in working memory\footnote{If not,
use more reducers or split the files.} and can be analyzed on distinct
compute nodes, each employing another layer of parallelization in
looping through Poisson regression for each token. This scheme is able
to take advantage of Hadoop for fast tokenization of distributed data,
and of
high performance computing architecture (much faster than, say, a
virtual AWS instance) for each regression. It is a model that should
work well for the many statisticians who have access to computing grids
designed for high throughput tasks more traditionally associated with
physics or chemistry.
\end{appendix}

%





\printaddresses
\end{document}